\documentclass[aps,pra,twocolumn,superscriptaddress,longbibliography, 10pt]{revtex4-2}
\usepackage{silence}
\usepackage{graphics,graphicx, array, afterpage}
\usepackage{qcircuit,amsmath}
\usepackage{amsfonts}
\usepackage{parskip}
\usepackage{grffile}
\usepackage[export]{adjustbox}
\usepackage{lineno}
\usepackage{booktabs}
\usepackage{tabularx}
\usepackage{array}
\usepackage{braket}
\usepackage{multirow}
\usepackage[latin1]{inputenc}
\usepackage[caption=false]{subfig}
\usepackage{amsthm}
\usepackage{bm}
\usepackage{layout}
\usepackage{float}
\usepackage{slashed}
\usepackage{amssymb}%
\usepackage[margin=0.75in]{geometry}
\usepackage{color}
\usepackage{soul}
\usepackage{mathtools}
\DeclarePairedDelimiter{\ceil}{\lceil}{\rceil}
\usepackage[colorlinks=true,citecolor=blue]{hyperref}
\usepackage[capitalise,compress]{cleveref}
\newcolumntype{C}{>{$}c<{$}}
\newcolumntype{R}{>{$}r<{$}}
\newcolumntype{L}{>{$}l<{$}}
\newcolumntype{Y}{>{\centering\arraybackslash}X}

\newcommand{\Yb}{$^{171}{\rm{Yb}}^{+} $}
\newcommand{\avg}[1]{\left \langle #1 \right\rangle}

\usepackage{xcolor}

\renewcommand{\epsilon}{\varepsilon}

\newcounter{para}

\makeatletter
\newcommand*\bigcdot{\mathpalette\bigcdot@{.5}}
\newcommand*\bigcdot@[2]{\mathbin{\vcenter{\hbox{\scalebox{#2}{$\m@th#1\bullet$}}}}}

\makeatother

\theoremstyle{definition}

\usepackage[english]{babel}
\makeatletter
\let\ORIbbl@fixname\bbl@fixname
\def\bbl@fixname#1{%
  \@ifundefined{languagealias@\expandafter\string#1}
    {\ORIbbl@fixname#1}
    {\edef\languagename{\@nameuse{languagealias@#1}}}%
}
\newcommand{\definelanguagealias}[2]{%
  \@namedef{languagealias@#1}{#2}%
}
\makeatother

\definelanguagealias{en}{english}
\definelanguagealias{EN}{english}

\newcommand{\GHZ}[2]{\text{GHZ}_{#1}^{#2}}

\begin{document}

\title{Demonstration of Shor encoding on a trapped-ion quantum computer}
\title{Demonstration of Shor encoding on a trapped-ion quantum computer}
\author{Nhung H. Nguyen}
\affiliation{Joint Quantum Institute and Department of Physics, University of Maryland, College Park, MD 20742, USA}
\author{Muyuan Li}
\affiliation{Departments of Electrical and Computer Engineering, Chemistry,
and Physics, Duke University, Durham, NC 27708, USA}
\author{Alaina M. Green}
\affiliation{Joint Quantum Institute and Department of Physics, University of Maryland, College Park, MD 20742, USA}
\author{Cinthia Huerta Alderete}
\affiliation{Joint Quantum Institute and Department of Physics, University of Maryland, College Park, MD 20742, USA}
\author{Yingyue Zhu}
\affiliation{Joint Quantum Institute and Department of Physics, University of Maryland, College Park, MD 20742, USA}
\author{Daiwei Zhu}
\affiliation{Joint Quantum Institute and Department of Physics, University of Maryland, College Park, MD 20742, USA}
\author{Kenneth R. Brown}
\affiliation{Departments of Electrical and Computer Engineering, Chemistry,
and Physics, Duke University, Durham, NC 27708, USA}
\author{Norbert M. Linke}
\affiliation{Joint Quantum Institute and Department of Physics, University of Maryland, College Park, MD 20742, USA}

\date{\today}

\date{\today}
% insert suggested PACS numbers in braces on next line

%!TEX root = ./main.tex

\begin{abstract}
	Fault-tolerant quantum error correction (QEC) is crucial for unlocking the true power of quantum computers. QEC codes use multiple physical qubits to encode a logical qubit, which is protected against errors at the physical qubit level.  
	Here we use a trapped ion system to experimentally prepare $m$-qubit GHZ states  and sample the measurement results to construct $m\times m$ logical states of the $[[m^2,1,m]]$ Shor code, up to $m=7$. The synthetic logical fidelity shows how deeper encoding can compensate for additional gate errors in state preparation for larger logical states. However, the optimal code size depends on the physical error rate and we find that $m=5$ has the best performance in our system. We further realize the direct logical encoding of the $[[9,1,3]]$ Shor code on nine qubits in a thirteen-ion chain for comparison, with $98.8(1)\%$ and $98.5(1)\%$ fidelity for state $\ket\pm_L$, respectively.

\end{abstract}
\maketitle 

%%%%%%%%%%%%%%%%%%%%%%%%%%%%%%%%%%%%%%%%%%%%%%%%%%%%%%%%
% \tableofcontents

%%%%%%%%%

%%%%%%%%%
\section{Introduction}

Fault-tolerant logical qubit encoding and fault-tolerant operations are required for executing quantum algorithms of sufficient depth to solve relevant problems \cite{gidneyHowFactor20482019,vonburgQuantumComputingEnhanced2020,shawQuantumAlgorithmsSimulating2020}.
Fault-tolerant operations, such as state preparation, syndrome measurement, error correction, logical gates, and measurements are designed such that any physical-level error they introduce is corrected at the logical level \cite{nielsenQuantumComputationQuantum2011}. When the physical error rate is below a certain threshold, the logical error can be made arbitrarily small by concatenation, i.e. using multiple layers of encoding \cite{knillThresholdAccuracyQuantum1996}, or taking advantage of natural robustness within the system \cite{kitaevFaulttolerantQuantumComputation2003}. The optimal method for fault-tolerant quantum computation is unknown and current methods offer trade-offs between encoding rate, threshold  \cite{KovalevPryadkoPRA2013,LiQCE2020}, and the number of available fault-tolerant gates \cite{Kesselring2018boundariestwist,GutierrezPRA2019,KrishnaPRX2019}. The same is true for near-term quantum error correction where only a limited amount of protection from physical-level errors will be available.

The Shor code \cite{shorSchemeReducingDecoherence1995} 
protects against all physical single-qubit Pauli errors. While the canonical $[[9,1,3]]$ code is based on triple modular redundancy, larger $[[m^2,1,m]]$ Shor codes can be generated using $m$-modular redundancy, where $m$ is the number of physical qubits in each module. 
The Shor code, together with the rotated surface code \cite{tomitaLowdistanceSurfaceCodes2014} and the Bacon-Shor subsystem code \cite{baconOperatorQuantumErrorcorrecting2006}, is an example of a compass code \cite{li2DCompassCodes2019}. The surface code has high memory and circuit-level thresholds, and treats phase- and bit-flip errors equivalently \cite{raussendorfFaultTolerantQuantumComputation2007,fowlerPracticalClassicalProcessing2012,stephensFaulttolerantThresholdsQuantum2014}. The Bacon-Shor code has no asymptotic threshold with $m$ for either $X$ or $Z$ errors \cite{nappOptimalBaconShorCodes2012}. The Shor code on the other hand has a memory threshold of 50\% for $Z$ errors and no threshold for $X$ errors as $m$ increases. In practice, this means that, for any physical error rate, there is an optimal size for the Shor and Bacon-Shor codes \cite{nappOptimalBaconShorCodes2012}.
These optimal codes can then be concatenated in a modular fashion to further improve performance \cite{aliferisSubsystemFaultTolerance2007}. Theoretical investigations comparing the 17-qubit rotated surface code \cite{tomitaLowdistanceSurfaceCodes2014} to a compass code on a 3$\times$3 qubit lattice find the latter to have much better performance in a realistic ion trap error model \cite{debroyLogicalPerformanceQubit2020}. In this paper, we find the optimal size $m$ of the Shor code that can be implemented on a particular trapped-ion quantum computer and investigate how measurements on a few qubits can predict the performance of larger systems.

Trapped ions are a promising platform for realizing a large-scale fault-tolerant quantum computer due to their long coherence time \cite{wangSinglequbitQuantumMemory2017}, high connectivity \cite{Linke3305,wrightBenchmarking11qubitQuantum2019}, high-fidelity single- and two-qubit gates \cite{ballanceHighFidelityQuantumLogic2016,gaeblerHighFidelityUniversalGate2016a}, and scalable architectures \cite{,pinoDemonstrationQCCDTrappedion2020a,monroeLargescaleModularQuantumcomputer2014}. Also, different components needed for fault-tolerance have been successfully demonstrated on trapped ions, such as logical state preparation \cite{niggExperimentalQuantumComputations2014,eganFaultTolerantOperationQuantum2021}, single-qubit logical operations \cite{niggExperimentalQuantumComputations2014,fluhmannEncodingQubitTrappedion2019,eganFaultTolerantOperationQuantum2021}, quantum error-detection with stabilizer readout \cite{linkeFaulttolerantQuantumError2017,eganFaultTolerantOperationQuantum2021}, magic state preparation \cite{eganFaultTolerantOperationQuantum2021} and multiple rounds of feedback-correction \cite{negnevitskyRepeatedMultiqubitReadout2018a}. Here we prepare $m$-qubit GHZ states on a trapped-ion system and extrapolate the logical error rate classically in order to emulate state preparation and measurement of an $[[m^2,1,m]]$ Shor code, where $m=3,4,5,6,7$. This emulation yields the optimal code size for our current system. We then compare the emulated $m=3$ results to the full $3\times3$ code state preparation on nine physical qubits.

The structure of the paper is as follows. In \cref{sec:bs-review} we review the Shor code and describe the methods used to study $[[m^2,1,m]]$ codes. In \cref{sec:trappedion} we outline the experimental setup. In \cref{sec:scaling} we present the experimental results on scaling. In \cref{sec:913} we demonstrate the logical basis state preparation of $m=3$ with 9 qubits. Lastly in \cref{sec:dis} we discuss the implications of these results for realizing fault-tolerant quantum computing. 

%-----------------------
\section{The Shor code}\label{sec:bs-review}
 
 An $[[m^2,1,m]]$ Shor code uses $m\times m$ physical qubits to encode a single logical qubit with distance $m$, i.e. any two orthogonal logical states differ by at least $m$ bit- or phase-flips. It is constructed from the concatenation of an $m$-bit repetition code that corrects X errors with an $m$-bit repetition code that corrects Z errors \cite{shorSchemeReducingDecoherence1995}.  Since all Pauli errors can be described as combinations of $Z$ and $X$ errors, measuring the stabilizers returns one of the potential syndromes, which give the location and type of the physical errors.  These can then be remedied by applying suitable $X$ and/or $Z$ correction operations. For the $[[9,1,3]]$ Shor code only a single physical error can be diagnosed unambiguously, since it has distance 3.

\begin{figure}[t]
\centering
\subfloat[\label{subfig:9plus}]{
\begin{tabular}[c]{l c}

\Qcircuit @C=0.7em @R=0.2em @!R {
% qubit 1
& |0\rangle & & \gate{H} & \ctrl{1} & \qw & \qw \\
% qubit 2
& |0\rangle & & \qw & \targ & \ctrl{1} & \qw \\
% qubit 3
& |0\rangle & & \qw & \qw & \targ & \qw \\
% qubit 4
& |0\rangle & & \gate{H} & \ctrl{1} & \qw & \qw\\
% qubit 5
& |0\rangle & & \qw & \targ & \ctrl{1} & \qw \\
% qubit 6
& |0\rangle & & \qw & \qw & \targ & \qw \\
% qubit 7
& |0\rangle & & \gate{H} & \ctrl{1} & \qw & \qw \\
% qubit 8
& |0\rangle & & \qw & \targ & \ctrl{1} & \qw\\
% qubit 9
& |0\rangle & & \qw & \qw & \targ & \qw \\
} \\
\end{tabular}
% \caption{Circuit to prepare $\ket{+}_L$}
}
\quad
\subfloat[\label{subfig:9minus}]{
\begin{tabular}[c]{l c}

\Qcircuit @C=0.7em @R=0.2em @!R {
% qubit 1
& |0\rangle & & \gate{H} & \ctrl{1} & \qw & \gate{Z} &\qw  \\
% qubit 2
& |0\rangle & & \qw & \targ & \ctrl{1} & \qw &\qw\\
% qubit 3
& |0\rangle & & \qw & \qw & \targ & \qw & \qw\\
% qubit 4
& |0\rangle & & \gate{H} & \ctrl{1} & \qw& \gate{Z}  & \qw \\
% qubit 5
& |0\rangle & & \qw & \targ & \ctrl{1} & \qw & \qw\\
% qubit 6
& |0\rangle & & \qw & \qw & \targ & \qw & \qw\\
% qubit 7
& |0\rangle & & \gate{H} & \ctrl{1} & \qw  &\gate{Z}  & \qw \\
% qubit 8
& |0\rangle & & \qw & \targ & \ctrl{1} & \qw & \qw\\
% qubit 9
& |0\rangle & & \qw & \qw & \targ & \qw & \qw\\
} \\
\end{tabular}
% \caption{Circuit to prepare $\ket{-}_L$}
}
\caption{Circuits for fault-tolerant preparation of logical states (a) $\ket+_L$ and (b) $\ket-_L$ of the $[[m^2,1,m]]$ Shor code for $m=3$. The circuit separates into $m$ groups, each preparing a $\ket{\GHZ m\pm}$ state. 
}\label{fig:FTShor3}. 
\end{figure}
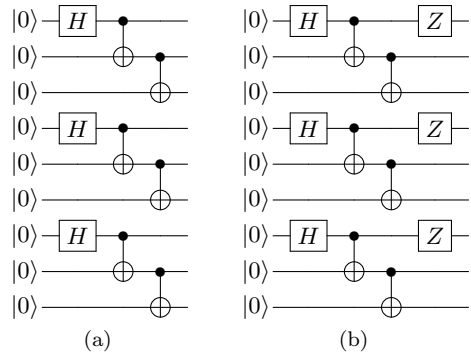

State preparation starts by fault-tolerantly creating a logical basis state, followed by fault-tolerant logical gates to generate a desired logical state $\ket{\psi}_L$. For an $[[m^2,1,m]]$ Shor code, the logical basis is given by $\ket{\pm}_L = \ket{\GHZ m\pm}^{\otimes m}$, where $\ket{\GHZ m\pm}=\frac1{\sqrt{2^m}}(\ket{0}^{\otimes m}\pm\ket{1}^{\otimes m})$. Since the $\ket{\pm}_L$ are product states of $\ket{\GHZ m\pm}$, we can prepare and measure many copies of a single $\ket{\GHZ m\pm}$ and randomly sample from these copies to artificially construct results corresponding to an $m\times m$ logical state. For example, with $m=3$, the logical states are $\ket{+}_L=\frac{1}{2\sqrt{2}}(\ket{000}+\ket{111})^{\otimes 3} = \bigotimes_{i=1,2,3} \ket{\GHZ3+}_i$ and $\ket{-}_L=\frac{1}{2\sqrt{2}}(\ket{000}-\ket{111})^{\otimes 3} = \bigotimes_{i=1,2,3} \ket{\GHZ3-}_i$. 
The circuit for encoding the $\ket{+}_L$ separates into three independent sub-circuits for creating three $3$-qubit GHZ states, see \cref{fig:FTShor3}. This is also the case for state preparation of the Bacon-Shor subsystem code \cite{eganFaultTolerantOperationQuantum2021}. This up-sampling allows us to study an $m\times m$-qubit Shor code with only $m$ qubits. However, some physical errors that come with larger system sizes such as cross-talk and others \cite{cetinaQuantumGatesIndividuallyAddressed2020} are underestimated.

 For a $[[9,1,3]]$ Shor code, there are eight stabilizers, six $Z$ stabilizers, which detect $X$ errors, and two $X$ stabilizers, which detect $Z$ errors \cite{debroyLogicalPerformanceQubit2020}. Therefore the code is better at detecting $X$ errors than $Z$ errors.
 To detect a single bit-flip within any GHZ sub-group, we measure the $Z$ stabilizers $ Z_jZ_{j+1}$ for the physical qubit index $j=1,2,4,5,6,7$. To detect a single phase-flip error, 
 we measure the $X$ stabilizers $X_1X_2X_3X_4X_5X_6$ and $X_4X_5X_6X_7X_8X_9$.
 These error detection measurements can be done in a non-destructive way by projecting the parity onto an ancilla qubit and measuring it without disturbing the code qubits \cite{liDirectMeasurementBaconShor2018}. In our experiment we directly perform the projective measurement on the physical qubits and perform the error detection or correction procedure in post-processing. While not realizing a full error correction scheme, we still observe how logical errors scale with the number of physical qubits and gates used for state preparation.

\section{Trapped ion setup}\label{sec:trappedion}

We carry out this experiment on a chain of trapped ions in a linear Paul trap. Two states in the hyperfine-split $^{2}$S$_{1/2}$ ground level of \Yb, $\ket{F=0,m_F=0}$ and $\ket{F=1,m_F=0}$, form the qubit. The ions are laser-cooled close to the motional ground state and initialized to $\ket{0}$ via optical pumping. Coherent operations are performed with two counter-propagating Raman laser beams, derived from a pulsed laser at $355$\:nm. The difference between relevant frequency components is stabilized to the energy splitting of the qubit. One of the Raman beams is split into an array of individual addressing beams, each of which is tightly focused onto exactly one ion, while the other is a global beam that illuminates the entire chain. We have frequency, amplitude, and phase control over each individual beam to selectively apply single-qubit and two-qubit gate operations. Detection is done via state-dependent fluorescence, where each ion is imaged onto one channel of a photo-multiplier tube array. Detailed performance of the system has been described elsewhere \cite{landsmanVerifiedQuantumInformation2019,figgattParallelEntanglingOperations2019}. For this work, we have extended the setup to operate on up to thirteen ions, at most nine of which act as qubits.

The native gate set consists of single-qubit rotations around any axis $\vec{n}_\phi$ in the $x$-$y$ plane, $R^j_\phi(\theta)=e^{-i\vec\sigma \cdot \vec{n}_\phi\theta/2}$,  rotations around the $z$-axis applied as classical phase shifts, $R^j_z(\theta)=e^{-i\sigma_z\theta/2}$, and two-qubit entangling gates $X_jX_k(\theta)=e^{i\sigma_x^j\sigma_x^k\theta}$ between any pair. These entangling gates are executed via spin-motion coupling based on the M\o lmer-S\o rensen scheme \cite{sorensenQuantumComputationIons1999,Solano99,milburnIonTrapQuantum2000}. We decouple the spin from the harmonic motion of all the modes by implementing a series of amplitude and frequency modulated pulses \cite{choiOptimalQuantumControl2014a,blumelEfficientStabilizedTwoqubit2021}. The fidelity for both single- and two-qubit gates is mainly limited by beam misalignment, beam-pointing instabilities, imperfect Stark-shift compensation and axial micromotion for all but the center ion. We do not have the ability to apply a quartic axial potential in order to space the ions equally. As a result, the alignment of the equally-spaced individual addressing beams worsens for larger numbers of ions in the trap. We mitigate this effect to some degree by using up to two additional ions at the each end of the chain. For five qubits, we trap seven ions; for seven qubits, we trap nine ions; for nine qubits, we trap thirteen ions. Trap imperfections also cause an unwanted axial radio-frequency (RF) field component that leads to axial micromotion. Therefore, in our setup, the single-qubit and two-qubit gate fidelities tend to decrease with the number of ions. The average fidelity for single-qubit gates (except $R_z$) in our experiment is $99.0(5)\%$  after correcting for state-preparation-and-measurement (SPAM) error. Typical fidelities for two-qubit gates are $99\%$ for a five-qubit system, $98.5\%$ for a seven-qubit system and $98\%$ for a nine-qubit system.

\section{Scaling of the Shor code}\label{sec:scaling}

\begin{figure}[t]
\centering
\includegraphics[width=0.48\textwidth]{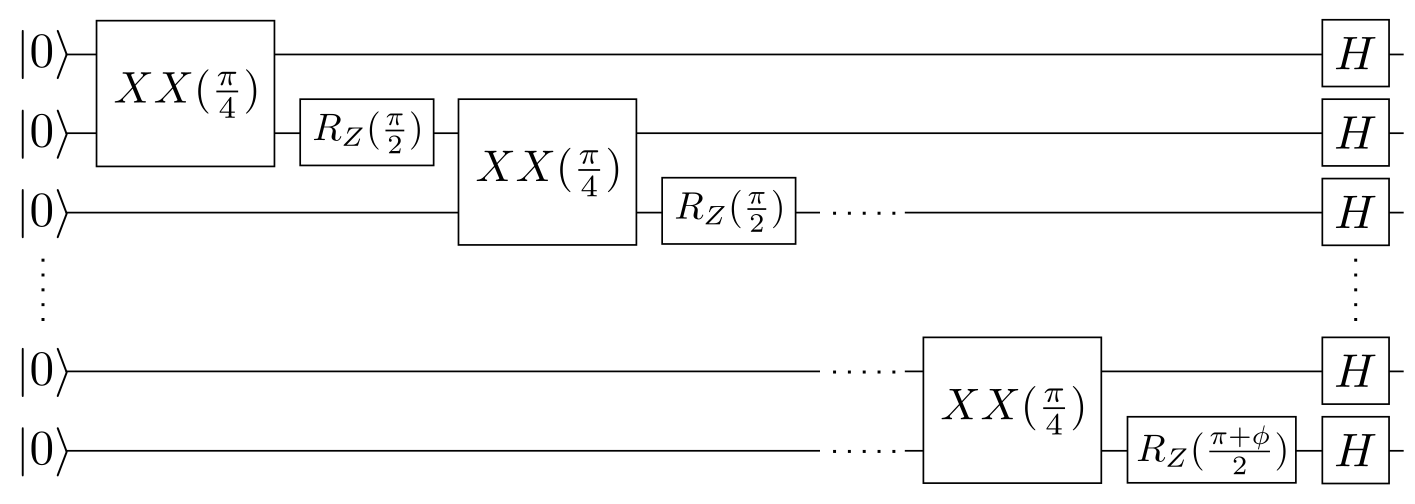}
\caption{The circuit to prepare a $\ket{\GHZ m\pm}$ on a trapped-ion quantum computer, with $\phi=0$ for $\ket{\GHZ m+}$ and $\phi=\pi$ for $\ket{\GHZ m-}$.}\label{fig:m-cat}
\end{figure}

\begin{figure}[t]
\centering
	\includegraphics[width=0.48\textwidth]{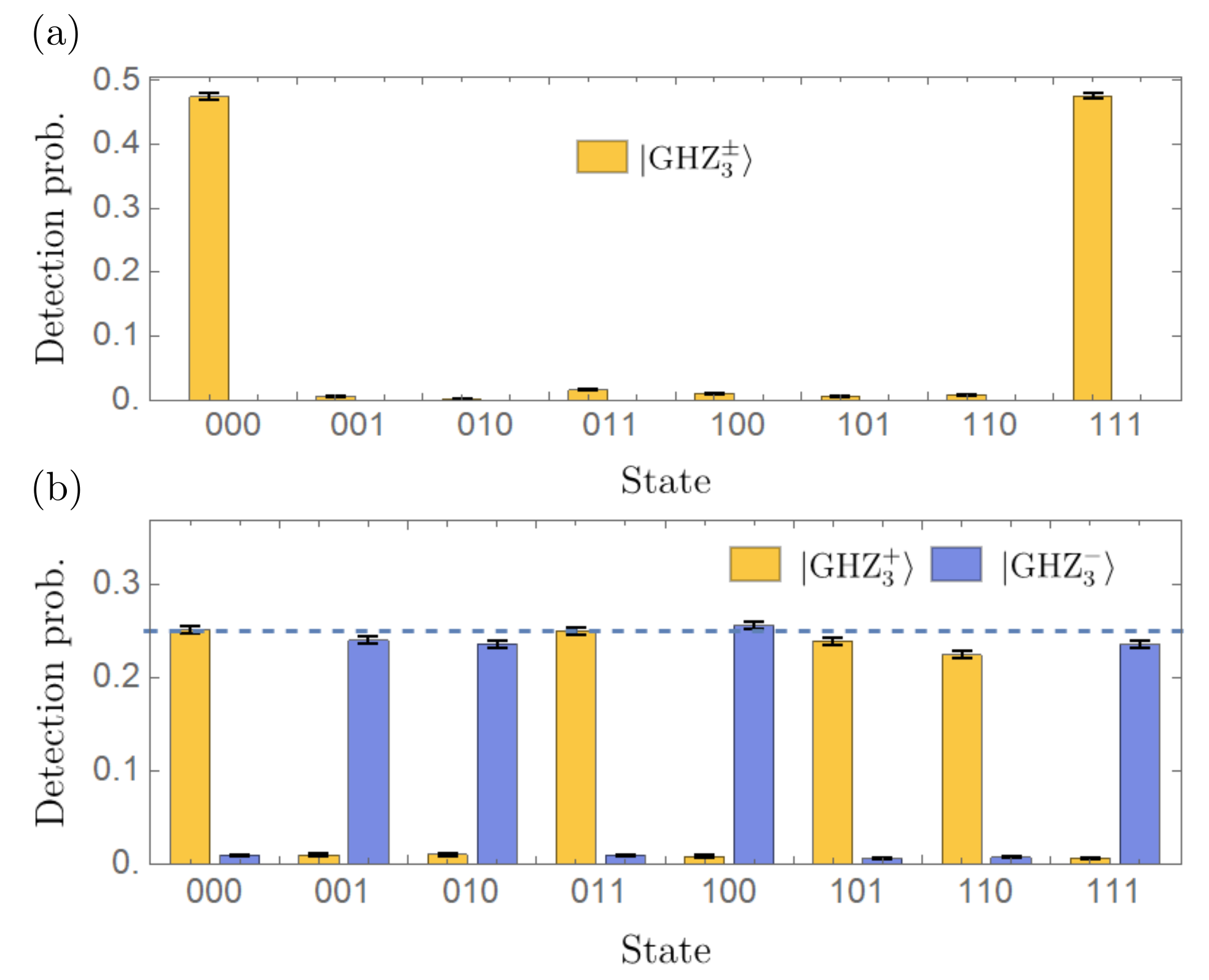}
	\caption{Measurement of $\ket{\GHZ3\pm}$ in the (a) $Z$ basis and (b) $X$ basis. In the Z basis, $\GHZ3\pm$ have the same measurement outcomes. The dashed line gives the ideal target population of $0.25$. }\label{fig:GHZ3Xbasis}
\end{figure}

\begin{figure}[bt]
\centering
	\includegraphics[width=0.45\textwidth]{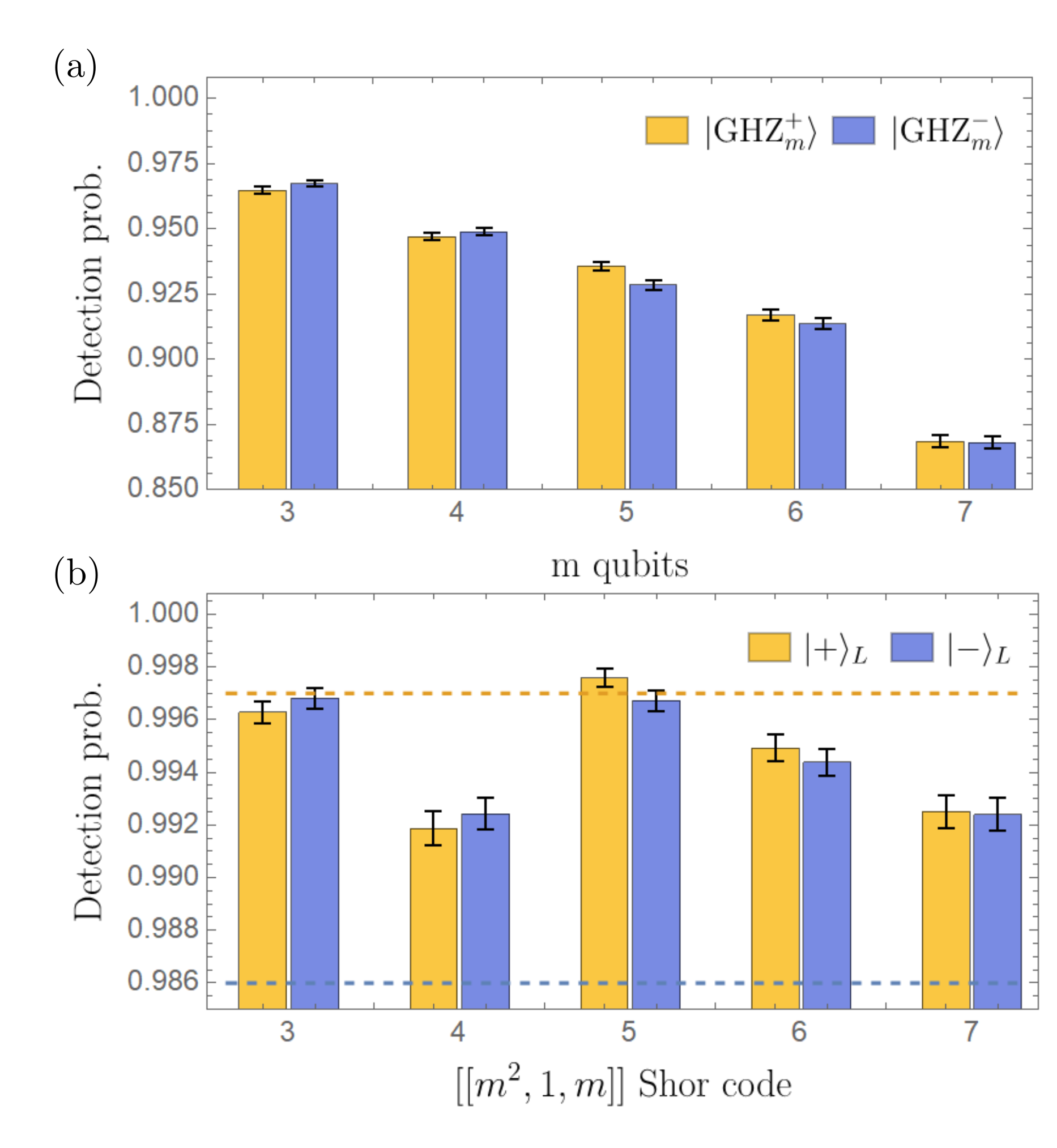}
	\caption{(a) $\ket{\GHZ m\pm}$ fidelity measured on $m$ trapped-ion qubits. (b) Up-sampled logical state fidelity of $[[m^2,1,m]]$ Shor codes after majority voting.  For even $m$, ties are assigned randomly. The dashed yellow (blue) line is the state preparation and measurement fidelity for state $\ket+$ ($\ket-$) of the physical qubit. Note that the vertical ranges for (a) and (b) are different. The increase in logical fidelity from $m=3$ to $m=5$ shows how deeper encoding can offer increased protection against physical errors.}\label{fig:scaling}
\end{figure}

We use the circuit in \cref{fig:m-cat} to create  $\ket{\GHZ m\pm}$ states using our native gate set. The results of measuring in the $Z$-basis for $m=3$ are shown in \cref{fig:GHZ3Xbasis}~(a). To measure in the $X$-basis, we apply $H^{\otimes m}$ before detection which creates an equal superposition of all even- and odd-parity computational states for $\ket{\GHZ m+}$ and $\ket{\GHZ m-}$, respectively, i.e. $\avg{X^{\otimes m}}=1$ and $-1$. The results for $m=3$ are shown in \cref{fig:GHZ3Xbasis}~(b). The probability of measuring the $\ket{\GHZ m \pm}$ in the $Z$ or $X$ basis, shown in \cref{fig:scaling}~(a), is given by summing the relevant measured state populations,
\begin{align}
	\mathcal{F}_z&=P_{00...0}+P_{11...1},\\
	\mathcal{F}_x^\pm &=\sum_{s \substack{\text{ even}\\\text{ odd}}} P_s.
\end{align}

 We sample a group of $m$ experimental shots from $\ket{\GHZ m\pm}$ to construct an artificial shot corresponding to the measurement of an $m\times m$ logical state $\ket{\pm}_L$, which we read out by majority voting. For even $m$, ties are assigned randomly. Repeating this $N/m$ times, where $N$ is the total number of experimental repetitions, we arrive at the fidelities $\mathcal{F}_L^\pm$ for $m =3,4,5,6,7$ shown in \cref{fig:scaling}~(b) and \cref{tab:scaling}. 

For large $N$ the up-sampled logical fidelities are given by 
\begin{align}
	 \mathcal{F}_L^\pm=\sum_{k=(m+1)/2}^m \binom{m}{k}(\mathcal{F}_x^\pm)^k(1-\mathcal{F}_x^\pm)^{m-k} \label{eq:logfid}
\end{align}

for odd $m$, and
\begin{align}
	 \mathcal{F}_L^\pm
	 &=\sum_{k=(m+2)/2}^m \binom{m}{k}(\mathcal{F}_x^\pm)^k(1-\mathcal{F}_x^\pm)^{m-k}\nonumber\\
	 &+\frac12 \binom{m}{m/2}(\mathcal{F}_x^\pm)^{m/2}(1-\mathcal{F}_x^\pm)^{m/2}  \label{eq:logfideven2}
\end{align}
for even $m$, due to the random assignment of ties. 

\begin{figure}[ht]
	\includegraphics[width=0.45\textwidth]{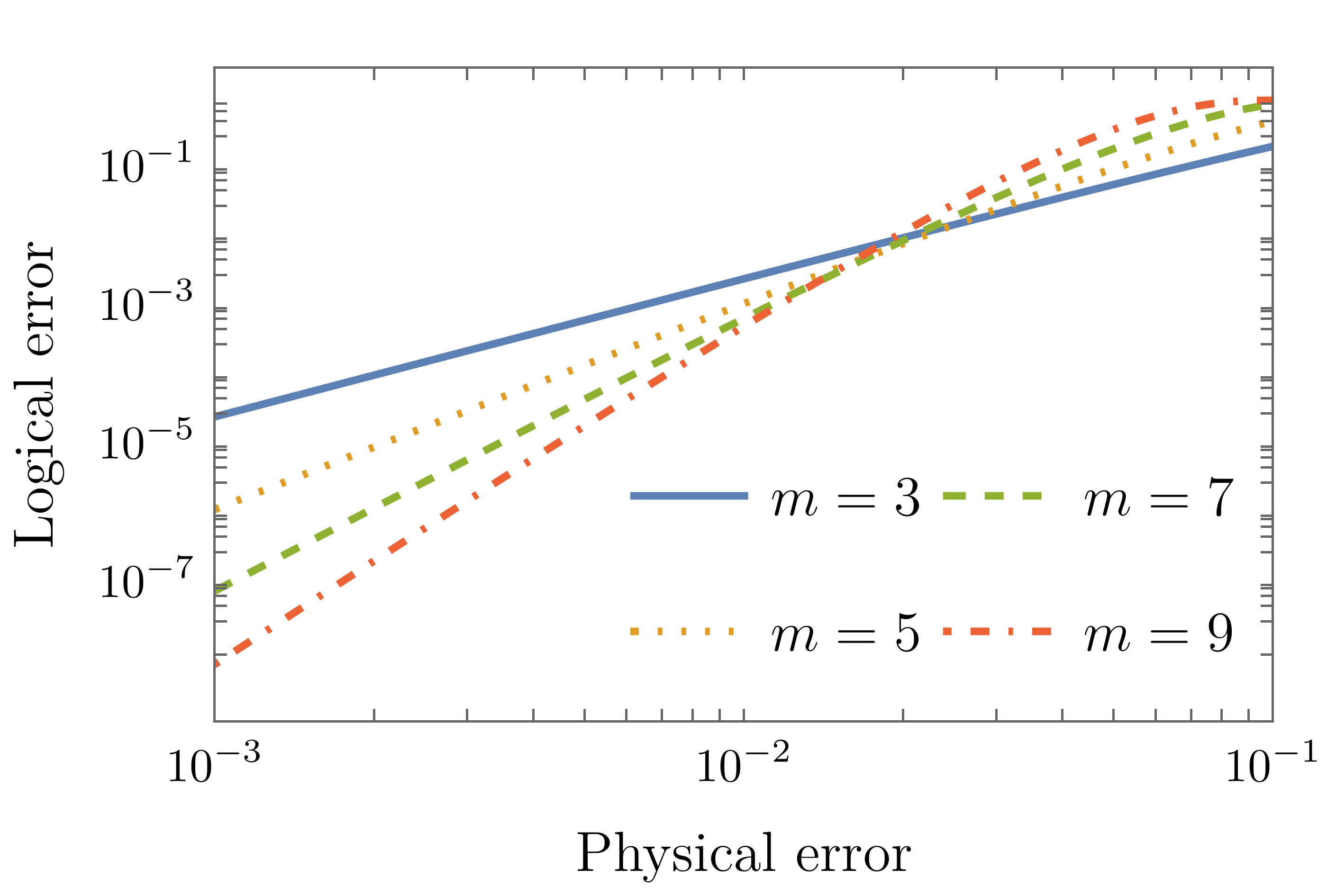}
	\caption{Scaling of the $[[m^2,1,m]]$ Shor code given by a simple depolarizing error model,  \cref{eq:logfid2}. } \label{fig:logfid2}
\end{figure}

Assuming depolarizing errors dominate, the physical qubit fidelity is roughly $f=\mathcal{F}_x^\pm/m$ \cite{nappOptimalBaconShorCodes2012} and therefore the logical fidelity is
\begin{align}
	 \mathcal{F}_L^\pm=\sum_{k=\ceil{m/2}}^m \binom{m}{k}(mf)^k(1-mf)^{m-k}. \label{eq:logfid2}
\end{align}
\cref{fig:logfid2} plots the dependence of the logical error rate, which is $1-\mathcal{F}_L^\pm$, on the physical error rate, which is $1-f$, for different code sizes $m =3,5,7,9$ as given by \cref{eq:logfid2}. There is a cross-over point in the physical error rate where deeper encoding compensates for the larger number of gate errors that can arise when preparing larger GHZ states. The experimental results presented in \cref{tab:scaling} and \cref{fig:scaling} follow the estimated fidelity given by \cref{eq:logfid,eq:logfideven2}.

 Although the fidelity to prepare five-qubit GHZ states is lower than that of the three-qubit GHZ states, the up-sampled logical states for the $[[25,1,5]]$ code has a higher fidelity than that for the up-sampled $[[9,1,3]]$ code after majority voting. This hints at the onset of fault-tolerance, since it demonstrates that deeper encoding can compensate for the increase in physical errors caused by employing more qubits and gates, leading to a lower logical error than a shallower code. This increase is not replicated when going to $[[49,1,7]]$, which shows that the state preparation errors have increased substantially as seen in the drop in the fidelity of $\GHZ7\pm$ (\cref{fig:scaling}). The random assignment of ties leads to a lower probability for $m=4, 6$ in \cref{fig:scaling}~(b).

\begin{table}[bt]
\caption{ Fidelity of state preparation and measurement for $\ket{\GHZ m\pm}$ (Measure) and logical states of $[[m^2,1,m]]$ Shor codes constructed by up-sampling with majority voting (Majority vote). Data are taken with $N=20000 $~shots. The uncertainty is given by the standard deviation of the binomial distribution $\sqrt{\mathcal{F}(1-\mathcal{F})/N}$.}\label{tab:scaling}. 
\begin{tabularx}{0.48\textwidth}{l C Y Y Y Y Y}
\toprule
  \multirow{2}{*}{$m$} & \multirow{2}{*}{Prep.} & \multicolumn{1}{c}{Z Meas.} & \multicolumn{2}{c}{X Meas.}     &    \multicolumn{2}{c}{Majority vote}                  \\
				  	 &	  			  &	 $\mathcal{F}_z$  &  $\mathcal{F}_x^+$  & $\mathcal{F}_x^-$       & $\mathcal{F}_L^+$    &   $\mathcal{F}_L^-$  \\ \midrule
 \multirow{2}{*}{3} &+ &	\multirow{2}{*}{0.951(1)}	& 0.965(1) 	& 0.035(1) 	& 0.9963(4) & 0.0037(1)  \\ 
 					&- &  					& 0.033(1)     & 0.967(1) 	& 0.0032(1) & 0.9968(4) \\ 
 \multirow{2}{*}{4} &+ &	\multirow{2}{*}{0.917(2)} & 0.947(2)     & 0.053(1) 	& 0.9919(6) & 0.0081(1) \\ 
 					&- &  					& 0.051(1) 	& 0.949(2) 	& 0.0076(1)  & 0.9924(6)\\ 
\multirow{2}{*}{5}  &+ &	\multirow{2}{*}{0.882(2)} & 0.936(2) 	& 0.064(1) 	& 0.9976(3) & 0.0024(1) \\
 					&- &  					& 0.072(1) 	& 0.928(2)		& 0.0033(1) & 0.9967(4)\\
\multirow{2}{*}{6}  &+ &	\multirow{2}{*}{0.806(2)} & 0.917(2) 	& 0.083(1)  	& 0.9949(5) &  0.0051(1) \\
 					&- &  					& 0.086(1) 	& 0.914(2)		& 0.0056(1)  & 0.9944(5) \\
\multirow{2}{*}{7}  &+ &	\multirow{2}{*}{0.723(2)} & 0.869(2) 	& 0.131(1) 	& 0.9925(6) & 0.0075(1) \\
 					&- &  					& 0.132(1) 	& 0.868(2)		& 0.0076(1) & 0.9924(6) \\\botrule
\end{tabularx}
\end{table}

\begin{table}[t]
\caption{ Fidelity of logical states of $[[m^2,1,m]]$ after discarding non-unanimous results (Error detect) and the success probability (Yield). Data are taken with $N=20000 $~shots.}\label{tab:discard} 
\begin{tabularx}{0.45\textwidth}{l C Y Y Y Y}
\toprule
  \multirow{2}{*}{$m$} & \multirow{2}{*}{Prepare} &\multicolumn{2}{c}{Error detect}     &   \multirow{2}{*}{Yield}                  \\
				   &	  &  $+$  & $-$       &    \\ \midrule
 \multirow{2}{*}{3}&+ & 0.99995(1) 	& 0.00005(1) 	& 0.898(4)  \\ 
 					&-  & 0.00003(1)     & 0.99997(1) 	& 0.905(4) \\ 
 \multirow{2}{*}{4}&+ & 0.99999(1)     & 0.00001(1) 	& 0.804(6) \\ 
 					&-  & 0.000009(1) 	& 0.999991(1) 	  & 0.810(6)\\ 
\multirow{2}{*}{5} &+ & 0.999998(1) 	& 0.000002(1) 	& 0.717(3)  \\
 					&-  & 0.000003(1) 	& 0.999997(1)	 & 0.690(4)\\
\multirow{2}{*}{6} &+ &0.9999996(2) 	& 0.0000004(1)  & 0.594(5)  \\
 					&-  & 0.0000010(1) 	& 0.9999990(1)	  & 0.581(5) \\
\multirow{2}{*}{7} &+ & 0.999997(1) 	& 0.000003(1)  	& 0.373(6)  \\
 					&-  & 0.000002(1)  	& 0.999998(1)	& 0.371(6) \\\botrule
\end{tabularx}
\end{table}
The additional errors in the logical state preparation and measurement (SPAM) process mainly come from an increase in single- and two-qubit gate errors for longer ion chains as discussed in \cref{sec:trappedion},
and read-out errors because of cross-talk between photo-multiplier tube channels, i.e. physical SPAM errors. Physical readout cross-talk accounts for $1-5\%$ infidelity in the $Z$ -measurements, depending on $m$. The rest is from two-qubit gates, which corresponds to an average of  $0.9\%$ error per gate for $m=3$, $1.3\%$ for $m=4$, $1.6\%$ for $m=5$, $1.7\%$ for $m=6$ and $2.2\%$ for $m=7$. These errors come from the increased beam mismatch discussed in \cref{sec:trappedion}. 

The Shor code can also be used as an error detection code, where non-unanimous votes are discarded rather than corrected, which leads to a finite yield. The fidelity and yield of the logical states after this procedure are presented in \cref{tab:discard}. The fidelities are higher than for the correction scheme since all $m$ qubits have to flip for a logical error to occur. For large $N$, the yield can be estimated by $(\mathcal{F}_x^\pm)^m+(1-\mathcal{F}_x^\pm)^m$. The optimal code size has increased to $6$, which is a valid size since ties don't play a role in the detection code.  

Using \cref{eq:logfid} we can estimate the minimal $\ket{\GHZ m\pm}$ fidelities needed in order for the up-sampled $m\times m$-qubit logical state to have the same fidelity as that of the $3\times3$-qubit logical state. For $m=5$, it is $0.93$; for $m=7$ qubits, it is $0.895$, which translates to an average infidelity of $1.7\%$ per two-qubit gate. 

\section{Logical qubit encoding}\label{sec:913}

\begin{figure}[ht]
\centering
	\includegraphics[width=0.45\textwidth]{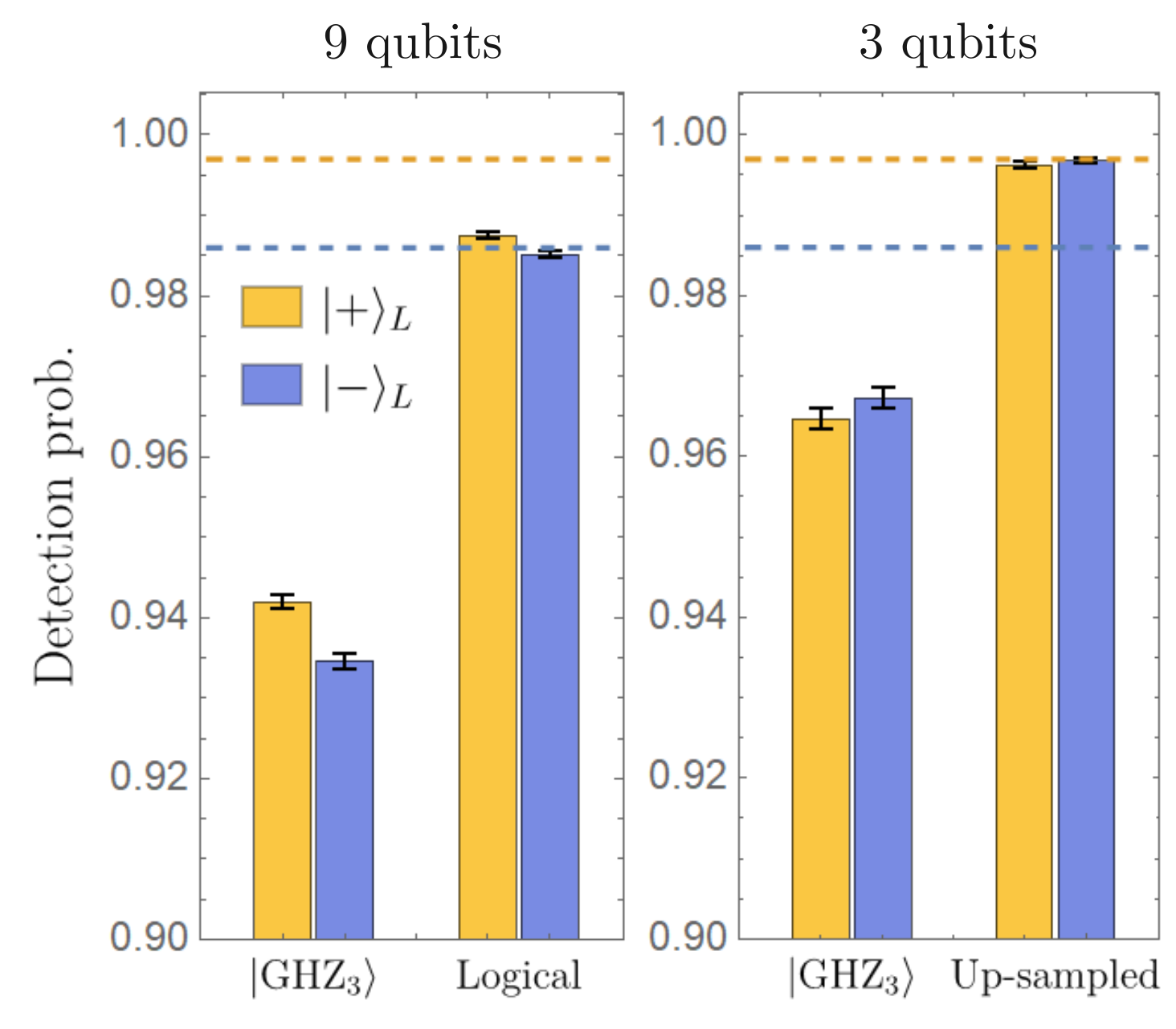}
\caption{A full $[[9,1,3]]$  Shor code logical state measurement with nine trapped-ion qubits (Left). We also show the average fidelity of the $\ket{\GHZ3\pm}$ states. For comparison, we show again the up-sampled results with three qubits from \cref{fig:scaling} (Right).  The dashed yellow (blue) line is the state preparation and measurement fidelity for state $\ket+$ ($\ket-$) of the physical qubit.}\label{fig:9qubit}
\end{figure}

We also perform the full $[[9,1,3]]$ encoding with nine qubits in a 13-ion chain. In this experiment we directly generate and read out logical states $\ket\pm_L$ of the $[[9,1,3]]$ Shor code. We also characterize the individual $\ket{\GHZ3\pm}$ states with measurements in the $X$-basis. The results are presented in \cref{tab:9qubit} and \cref{fig:9qubit}.

\begin{table}[bt]
\caption{Fidelity of state preparation and measurement for three sets of $\ket{\GHZ3\pm}$ labeled as $1,2,3$ and the logical state $\ket\pm_L$ of [[9,1,3]]] code on a thirteen-ion chain.} \label{tab:9qubit}
	\begin{tabularx}{0.4\textwidth}{c Y Y Y}
		\toprule
		\multicolumn{2}{r}{Prepare} &\multicolumn{2}{c}{Measure}\\
		 & & $+$ &  $-$ \\ \midrule
		\multirow{2}{*}{Logical}& $+$  & 0.988(1) & 0.012(1) \\
								& $-$   & 0.015(1) & 0.985(1) \\
		\multirow{2}{*}{$\ket{\GHZ3\pm}_1$}& $+$  & 0.942(1) & 0.058(1) \\
								& $-$   & 0.050(1) & 0.950(1) \\
		\multirow{2}{*}{$\ket{\GHZ3\pm}_2$}& $+$  & 0.942(1) & 0.058(1) \\
								& $-$   & 0.079(1) & 0.921(1) \\
		\multirow{2}{*}{$\ket{\GHZ3\pm}_3$}& $+$  & 0.942(1) & 0.058(1) \\
								& $-$   & 0.068(1) & 0.932(1) \\\botrule	
	\end{tabularx}
\end{table}

The fidelity of the logical states $\ket{\pm}_L$ is at the level of the physical state $\ket{-}$ and hence falls short of the average performance of the physical qubit $\ket{\pm}$. Using \cref{eq:logfid}, we estimate that the fidelity of each $\ket{\GHZ3+}$ state must be increased to $0.968$ in order to achieve the same logical fidelity as our physical qubit $\ket+$. It is worth noting that the fidelities of the $\ket{\GHZ3\pm}$ triplet on nine qubits are very similar to each other. This indicates a high level of uniformity among the qubits and gates.
 
 Notice that the fidelities of both the $\ket{\GHZ3\pm}$ states and logical $\ket{\pm}_L$ states are lower than their corresponding counterparts in the up-sampled experiment with three qubits in a chain of seven ions (see \cref{fig:scaling}). The additional errors reduce the fidelity compared to the up-sampled version by about $1\%$. 

% \par
\section{Discussion \& Outlook}\label{sec:dis}

Our preparation and sampling of $\ket{\GHZ m\pm}$ states to synthetically construct $m \times m$ logical Shor code states shows experimentally that deeper encoding can compensate for additional physical errors from logical state preparation. 
For our specific setup, the $\ket{\GHZ 5\pm}$ state projects the best logical fidelities. The increase in physical errors with larger $m$ we observe is due to hardware limitations, 
most of which can be solved by better engineering. For example, detection cross-talk can be eliminated by independent photon-detectors \cite{Slichter:17,crainHighspeedLowcrosstalkDetection2019,eganFaultTolerantOperationQuantum2021}.
The beam-ion alignment can be improved by traps with more control electrodes, such as micro-fabricated surface traps \cite{osti_1237003} or blade traps with more segments \cite{Pagano_2018}. Alternatively, near-perfect ion addressing can be achieved with integrated optics \cite{mehtaIntegratedOpticalMultiion2020,niffenegger2020} or beam steering using a micro-electro-mechanical system of mirrors \cite{wangHighFidelityTwoQubitGates2020}. Axial RF stray fields, and hence axial micromotion is also greatly reduced in precision-fabricated surface or 3D traps \cite{Pyka2014,decaroli2021design}.

The comparison of the emulated $9$-qubit Shor state to a direct preparation of the $9$-qubit Shor code states shows a $2\%$ decrease in the fidelity of individual $\ket{\GHZ m\pm}$ states and a $1\%$ decrease in the SPAM logical fidelity. This result points toward future work where parts of quantum error corrected codes can be used as benchmarks for system scalability and uniformity. Recent work has emphasized that physical errors are not uniform over the Pauli operators \cite{li2DCompassCodes2019,puriBiaspreservingGatesStabilized2020,ataidesXZZXSurfaceCode2020,tuckettFaulttolerantThresholdsSurface2020}. Although the $m\times m$ Shor code has no threshold for one type of Pauli error, it matches the classical threshold for the other Pauli error. This makes the Shor code an exciting choice for quantum memories with asymmetric errors. Further work on asymmetric Shor codes and how they interact with bias-preserving gates is warranted \cite{chamberlandBuildingFaulttolerantQuantum2020}.

\section*{Acknowledgments}
We are grateful to T. Yoder for helpful discussions and comments on the manuscript. We also thank S. Debnath, K. A. Landsman, C. Figgatt, and C. Monroe for for early contributions to this work.  Research at UMD was supported by the NSF, grant no. PHY-1430094,  to  the  PFC@JQI. A.M.G. is supported by a JQI Postdoctoral Fellowship. Research at Duke was supported by the Office of the Director of National Intelligence - Intelligence Advanced Research Projects Activity through an Army Research Office contract (W911NF-16-1-0082) and the Army Research Office (W911NF-21-1-0005). 
\bibliography{bacon-shor-9}

\pagebreak

\appendix

\end{document}